\documentclass[prd,aps,preprintnumbers,twocolumn]{revtex4}
\usepackage{epsfig}
\usepackage{amsmath}

\def\as{\ensuremath{\alpha_{s}}}
\def\a0{\alpha_0}

\def\vep{\varepsilon}
\def\f{{\rm f}}

\def \ep{\epsilon}

\def\bea {\begin{eqnarray}}
\def\eea {\end{eqnarray}}

\def\be {\begin{equation}}
\def\ee {\end{equation}}

\begin{document}

\preprint{YITP-SB-09-06}

\renewcommand{\thefigure}{\arabic{figure}}

\title{The Massive Soft Anomalous Dimension Matrix at Two Loops}

\author{Alexander Mitov, George Sterman, Ilmo Sung}

\affiliation{C.N.\ Yang Institute for Theoretical Physics,
Stony Brook University, Stony Brook, New York 11794--3840, USA}

\date{March 18, 2009}%\today}

\begin{abstract}
We study  two-loop anomalous  dimension matrices in QCD and related gauge theories for products of Wilson lines coupled at a point.  We verify by an analysis in Euclidean space that the contributions to these matrices from diagrams that link three massive Wilson lines do not vanish in general.   We show, however, that for two-to-two processes the two-loop anomalous dimension matrix is diagonal in the same color-exchange basis as the one-loop matrix for arbitrary masses at absolute threshold and for scattering at ninety degrees in the center of mass.   This result is important for applications of threshold resummation in heavy quark production.
\end{abstract}

\maketitle

The infrared structure of perturbative amplitudes is relevant for a
variety of hard-scattering processes, including the production of
heavy particles, whether charged under QCD or not, and of high-$p_T$
jets. Infrared enhancements in these amplitudes are typically
regularized by continuing to $D$ dimensions, where they appear as
poles in $\vep=2-D/2$. Infrared poles, of course, do not appear in
physical predictions for infrared safe quantities, in which they
cancel after an appropriate sum over final states.  Nevertheless,
the all-order structure of infrared poles is of interest for exact
fixed-order calculations \cite{QQbarfixed}, for resummations of
long-distance enhancements in cross sections \cite{kidonakisrv}  and
for supersymmetric gauge theories \cite{Alday:2008yw}.

In this paper, we study the dependence of infrared enhancements
at two loops on the masses of external partons.
Our analysis is relevant to
dimensionally-regularized  amplitudes for $2 \to n$
processes (labelled by {\bf f}),
\bea \label{f_1f_n} {\bf f}:
\quad f_1(p_1,r_1) + f_2(p_2,r_2) \to & \nonumber\\
&& \hspace{-30mm} f_3(p_3,r_3)  + \dots  f_{n+2}(p_{n+2},r_{n+2})  \, ,
\eea
where the $r_i$ represent color indices. At fixed, nonforward
angles, amplitudes of this type factorize into functions
representing short-distance (hard) dynamics, collinear dynamics of
external massless partons, and soft-gluon exchange between light and
heavy partons.   We can represent such a factorized amplitude as
\cite{matrixdim,TYS}
\begin{eqnarray}
\label{facamp}
\left |\, {\cal M}_\f
  \left(\beta_i,\frac{Q^2}{\mu^2},\as(\mu),\vep \right)\right \rangle
&=&\prod_i J^{[i]}\left(\as(\mu),\vep
\right)
\nonumber\\
&\ & \hspace{-45mm}
\times\, {\bf S}_\f\left( \beta_i,\frac{Q^2}{\mu^2},\as(\mu),
                     \vep \right)
\left| \, H_\f
  \left(\beta_i,\frac{p_j}{\mu},\frac{Q}{\mu},\as(\mu)\right) \right \rangle\, ,
  \end{eqnarray}
in the notation of Ref.\ \cite{catani96}, where the ket $|A(\rho,\vep)\rangle$
represents $\sum_{L=1}^C A_L(\rho,\vep)\
\left(c_L\right)_{\{r_i\}}$, with $\{c_L\}$ some basis of color
tensors linking the heavy and light parton lines at short distances.
The jet functions $J^{[i]}$ for incoming and outgoing lightlike
partons may be defined, for example, as the square root of the
corresponding elastic form factor \cite{TYS}. The soft matrix ${\bf
S}_\f$, which is the subject of study here, describes the infrared
behavior of color exchange between the external partons. It remains
after the factorization of collinear enhancements into the jet
functions for those external partons that are massless, and in
general mixes the color components found at short distance.   We do
not introduce jet functions for massive partons here.   The
resummation of collinear logarithms associated with high-$p_T$
massive partons can  be treated by factorization as well
\cite{MassiveFermions}, but this is not a goal of this paper.

The soft matrix is determined entirely by a matrix of anomalous
dimensions \cite{matrixdim}, through
\bea \label{expoS} &&
\hspace{-5mm} {\bf S}_\f \left(\beta_i,\as(\mu),\vep \right)
\nonumber\\
&&  =
{\rm P}~{\rm exp}\left[
\, -\; \int_{0}^{\mu} \frac{d\tilde{\mu}}{\tilde{\mu}}
{\bf \Gamma}_{S_\f} \left(\beta_i\cdot\beta_j ,
  \bar\as\left(\tilde{\mu}, \vep\right)\right) \right]\, ,
\end{eqnarray}
where the velocities $\beta_i$ are lightlike for light partons,
$\beta_l^2=0$, and may be scaled to unit length for heavy particles,
$\beta_h^2=1$. Integrals in the exponent are carried out in $D>4$,
using the $D$-dimensional running coupling,
$\bar\as\left(\tilde{\mu}, \vep\right)$. Equation (\ref{expoS}),
along with corresponding expressions for the jet functions,
determines the infrared pole structure to all orders in perturbation
theory for processes involving the wide-angle scattering of any
number of massless {\it and} massive partons 
\cite{matrixdim,TYS,Kidonakis:1997gm}.

The determination of the anomalous dimension matrix for an arbitrary
process with massless and massive partons is equivalent to the
renormalization of a set of color tensors that link the
corresponding product of Wilson lines at a point
\cite{Brandt:1981kf,Kidonakis:1997gm}.  Each Wilson line follows the
velocity $\beta_i$ of the corresponding parton, without recoil, from
this point to infinity, either from the initial state or into the
final  state.   These composite operators mix under renormalization
in general, leading to the matrix structure shown in Eq.\
(\ref{expoS}). The one-loop anomalous dimensions for gluons and for
both massless and massive quarks have been known for some time
\cite{Kidonakis:1997gm,DMFKS}. At two loops, the matrices for any $2
\to  n$ process with {\it massless} lines satisfy the relation
\cite{Aybat} \bea {\bf \Gamma}_{S_\f}^{(2)} (\beta_i) =
\frac{K}{2}\, {\bf \Gamma}_{S_\f}^{(1)}(\beta_i) \, , \label{Gamma2}
\eea with ${\bf \Gamma}_{S_\f}^{(i)}$ the coefficient of
$(\as/\pi)^i$, and with $K=C_A(67/18 -\zeta(2)) - 10T_Fn_F/9$. This
is exactly the relation satisfied by the expansion of the cusp
anomalous dimension \cite{cusp}, which generates the leading, double
poles in the elastic form factor \cite{Sudrefs,magnea90}.

\begin{figure}
{\epsfxsize=3 cm \epsffile{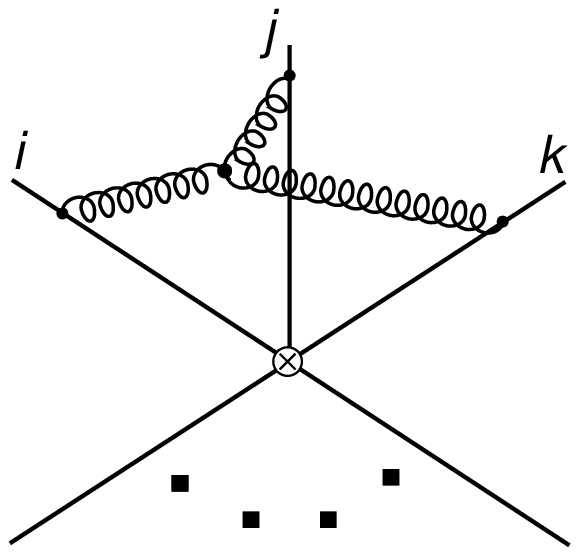} \quad \epsfxsize=3 cm \epsffile{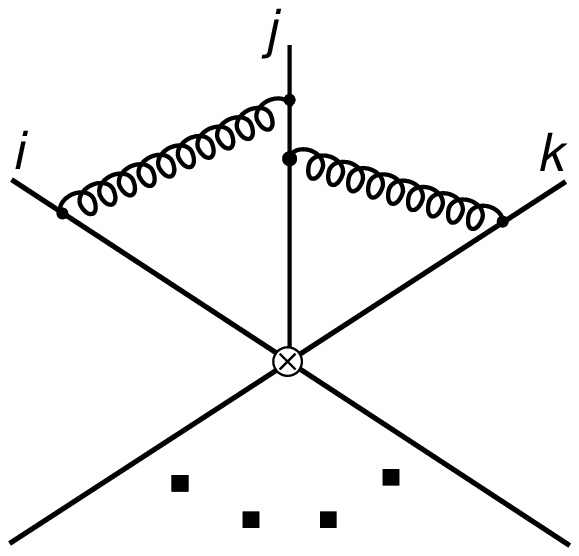} \\
\hbox{ \hskip 2.25 cm (a) \hskip 3.0 cm (b) }
\quad \epsfxsize=3 cm \epsffile{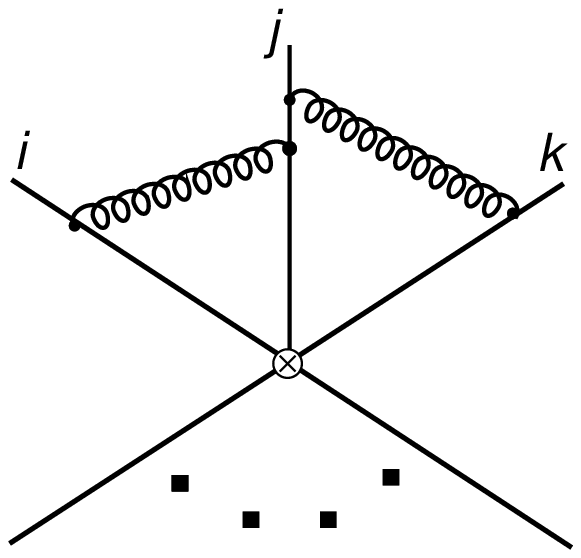} \quad \epsfxsize=3 cm \epsffile{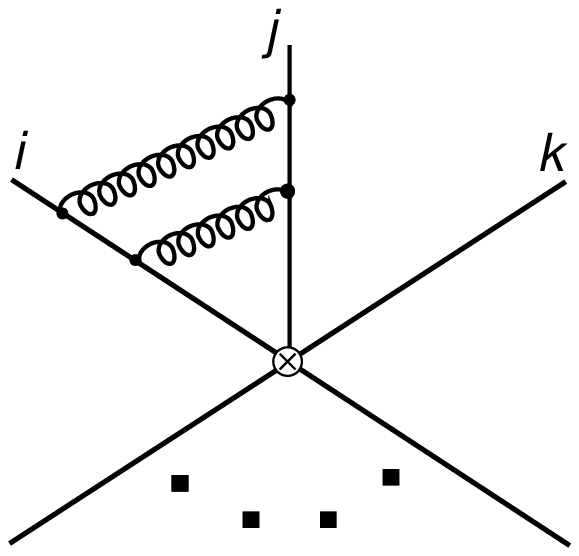}\\
\hbox{ \hskip 2.25 cm (c) \hskip 3.0 cm (d) }
\quad \epsfxsize=3 cm \epsffile{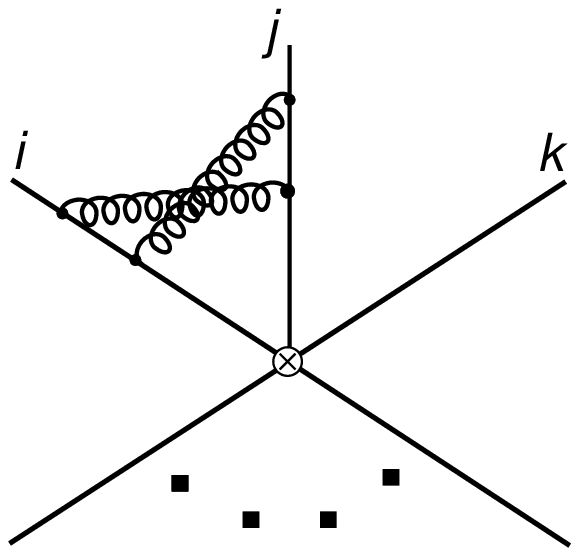} \quad \epsfxsize=3 cm \epsffile{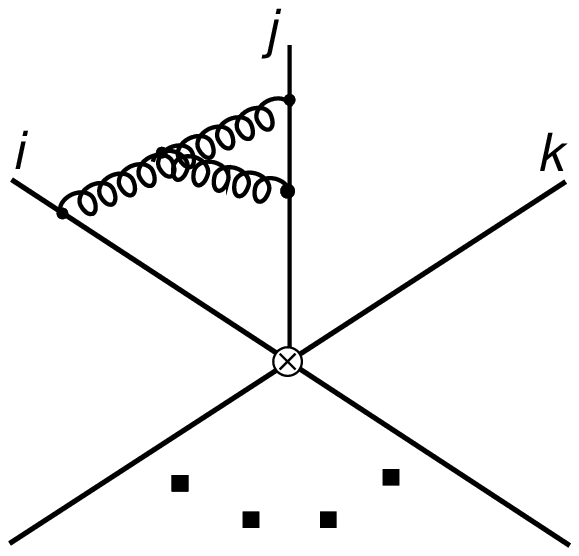}}\\
\hbox{ \hskip 2.25 cm (e) \hskip 3.0 cm (f) }
\caption{\label{3Efig} Diagrams whose ultraviolet poles
determine the soft anomalous dimension
at two loops.   The crossed vertex represents 
the point at which the Wilson lines are linked.    The straight lines represent
eikonal propagators.   Referring to the number of Wilson lines linked by
gluons, in the text we refer to these as
3E diagrams   (a-c) and
2E diagrams (d-f). \label{diagrams}}
\end{figure}

Examples of the diagrams involved in the calculation of the two-loop
anomalous dimension are shown in Fig.\ \ref{diagrams}.   
In momentum space, the propagators and vertices from Wilson lines
are given by eikonal expressions \cite{Kidonakis:1997gm}.
The
corresponding two-loop corrections to the anomalous dimensions are
found in the usual way \cite{Aybat} from the two-loop UV
single poles of these diagrams after one-loop renormalization.

The result (\ref{Gamma2}) for massless partons is a consequence of
the vanishing of the single poles of those two-loop ``3E" diagrams
in which color is  exchanged coherently between three eikonal lines
in the figures. The arguments of Ref.\ \cite{Aybat} do not, however,
generalize directly to massive Wilson lines, with velocity vectors
$\beta_i^2\ne 0$. While an analytic determination of
$\Gamma_S^{(2)}$ would, of course, be desirable, numerical
determination is also of interest, and is certainly adequate to
determine how far Eq.\ (\ref{Gamma2}) generalizes to the production
of massive particles.   
We provide the necessary analysis below, and
show that when the $\beta_i^2$ are nonzero, 
Eq.\ (\ref{Gamma2}) no longer holds.
A generalization of Eq.\ (\ref{Gamma2}), however, given by Eq.\ (\ref{Gamma2m}) below,
does holds for two-to-two processes  for special momentum configurations.

Much of our analysis will be carried out in position,  rather than
momentum  space. In the following, we will take every parton as
massive, and use the scale invariance of Wilson lines to set
$\beta_i^2=1$. Because we are calculating renormalization
constants, we can carry out our analysis in  Euclidean space.
Indeed, a numerical result in Euclidean space is adequate to
establish that the matrix does not follow Eq.\ (\ref{Gamma2}) in
Minkowski space. Otherwise, analytic continuation through Wick
rotation would imply that the same result would hold in Euclidean
space as well.

We begin with the diagram, Fig.\ \ref{diagrams}a, in which three
eikonal lines are coupled by gluons that are linked at a three-gluon
coupling  \cite{Aybat}. In the configuration space evaluation of
this diagram, we must integrate the position of the three-gluon
vertex over all space. The three propagators each have one end fixed
at this vertex and the other end fixed at a point $\lambda_i\beta_i$
along the $i$th Wilson line.   Each parameter $\lambda_i$ is
integrated from the composite vertex to infinity.   This diagram
vanishes in Minkowski space for massless Wilson lines \cite{Aybat}.

Suppressing color factors, we represent the 3E diagram Fig.\ \ref{diagrams}a as
\bea F^{(2)}_{3g}(\beta_I) = \int d^Dx \, \prod_{i=1}^3
\int_0^\infty d\lambda_i V(x,\beta_I)\, . \label{Gam3g}
\eea
Here $\beta_I=\{\beta_1,\beta_2,\beta_3\}$ denotes the set of three
massive velocities of the lines to which the gluons attach, while
the propagators and numerator factors of the integrand are given by
a sum over six terms, \bea V(x,\beta_I) = \sum_{i,j,k=1}^3 \ep_{ijk}
v_{ijk}(x,\beta_I)\, . \label{Vv} \eea Each of these terms involves
the derivative of one of the propagators, according to the usual
gauge theory rules for the three-vector coupling,
\bea v_{ijk}(x,\beta_I) &=& -i(g\mu^\vep)^4
\beta_i\cdot \beta_j\, \Delta(x-\lambda_j\beta_j)\, \Delta(x-\lambda_k\beta_k)\nonumber\\
&\ & \hspace{15mm} \times\, \beta_k\cdot \partial_x \Delta(x-\lambda_i\beta_i)
\, ,
\label{vDelta}
\eea
where $\Delta$ represents the position-space scalar propagator,
\bea \Delta(x-\lambda_i\beta_i) =
-~\frac{\Gamma(1-\vep)}{4\pi^{2-\vep}} \frac{1}{\left(
x-\lambda_i\beta_i\right)^{2(1-\vep)}}\, . \label{Deltadef}
\eea
We work in Feynman gauge.    The contribution of this (scaleless)
diagram to the anomalous dimension matrix is found from the residue
of its simple ultraviolet pole. We note that all diagrams found from
products of Wilson lines are scaleless overall, and are {\it
defined} by their renormalization constants \cite{Aybat}.

At fixed $x$, for massive eikonals the $\lambda$ integrals in Eq.\
(\ref{Gam3g}) are all finite in four dimensions.   After these
integrals are carried out, the $\beta_i$-dependence enters only
through the combination
\bea \zeta_i \equiv  \frac{\beta_i\cdot x}{\sqrt{x^2}}\, ,
\label{zetadef}
\eea
and we can write
\bea F^{(2)}_{3g}(\beta_I,\vep) = {\cal N}(\vep) \int d^Dx \,
\sum_{i,j,k=1}^3 \ep_{ijk} \gamma_{ijk}(\sqrt{x^2},\zeta_I,\vep)\, ,
\nonumber\\
\label{Gameps}
\eea
where ${\cal N}(\vep)$ absorbs overall factors that are finite in
the limit $\vep = 0$. To simplify our notation, in the following we
normalize $F_{3g}$ so that ${\cal N}(\vep) = 1$. We recall that we have used the
scale invariance of eikonal lines to set $\beta_I^2=1$, and that $I$
represents the set $i,j,k$. Each term $\gamma_{ijk}$ is now given by
\bea \gamma_{ijk}(\sqrt{x^2},\zeta_I,\vep) &=& \beta_i\cdot\beta_j\,
f(x,\beta_j,\vep)\, f(x,\beta_k,\vep)
\nonumber\\
&\ & \hspace{5mm} \times\, \beta_k\cdot\partial_x\, f(x,\beta_i,\vep)\, ,
\label{gammaf}
\eea
where the functions $f(x,\beta,\vep)$ are simply the integrals
of the $x$-dependent factors of the propagators,
\bea f(x,\beta,\vep) &=& \int_0^\infty d\lambda\ \frac{1}{\left(
x^2-2\lambda\beta\cdot x +\lambda^2 \right)^{1-\vep}}\, .
\eea
After a change of variables to $\lambda' \equiv \lambda/\sqrt{x^2}$,
the dependence on the variables $x^2$ and $\zeta_i$ factorizes,
\bea f(x,\beta_i,\vep) &=&
\frac{1}{\left(\sqrt{x^2}\right)^{1-2\vep}} \int_0^\infty d\lambda'\
\frac{1}{\left( 1-2\lambda'\zeta_i +\lambda'{}^2 \right)^{1-\vep}}
\nonumber\\
&\equiv& \frac{1}{\left(\sqrt{x^2}\right)^{1-2\vep}}\ g(\zeta_i,\vep)\, .
\label{fdefg}
\eea
For the full expression, we also need the gradient of this function,
which can be written as
\bea
\partial^i_x\, f(x,\beta,\vep) &=&
\frac{1}{\left(\sqrt{x^2}\right)^{2 - 2\vep}}\,
\bigg[ \, (2\vep-1) \frac{x^i}{\sqrt{x^2}}\,  g(\zeta,\vep)\
\nonumber\\
 &\ & \hspace{0mm}
+\  \left( -\ \frac{x^i\zeta}{\sqrt{x^2}} + \beta^i
\right)\, \frac{\partial g(\zeta,\vep)}{\partial \zeta} \, \bigg]\, .
\label{fgrad}
 \eea
We note that this derivative is necessary to
produce an overall $x^{-4}$ fall-off at infinity and a singularity at
$x=0$, corresponding to logarithmic infrared and ultraviolet
behaviors.

We next substitute the expressions for the functions $f$ in
(\ref{fdefg}) and their gradients (\ref{fgrad}) into Eq.\
(\ref{gammaf}) for the terms $\gamma_{ijk}$, to  find
\bea
\gamma_{ijk}(\sqrt{x^2},\zeta_I,\vep) &=& g(\zeta_j,\vep)
g(\zeta_k,\vep)\ \frac{\beta_i\cdot
\beta_j}{\left(\sqrt{x^2}\right)^{4-6\vep}}
\nonumber\\
&\ & \hspace{-38mm} \times \left[ (2\vep -1) \zeta_k g(\zeta_i,\vep)
+ \left( - \zeta_k\zeta_i + \beta_i\cdot \beta_k \right)
\frac{ \partial g(\zeta_i,\vep)}{\partial \zeta_i} \right].
\eea
In this expression, the first term in the square brackets is
symmetric in the pair $(i,j)$ and the third is symmetric in the pair
$(j,k)$. The full nonvanishing contribution to Eq.\ (\ref{Gameps})
is thus simply
\bea F^{(2)}_{3g}(\beta_I,\varepsilon) &=& -\ \int d^Dx \,
\sum_{i,j,k=1}^3 \ep_{ijk} \,  \zeta_k\zeta_i\, \frac{\beta_i\cdot
\beta_j}{\left(\sqrt{x^2}\right)^{4-6\vep}}
\nonumber\\
&\ & \hspace{5mm}
\times\ g(\zeta_j,\vep) g(\zeta_k,\vep)\
 \frac{ \partial g(\zeta_i,\vep)}{\partial \zeta_i}\,.
 \label{gggprime}
\eea
Using the freedom to reintroduce dependence on the $\sqrt{\beta_i^2}$ by
demanding scale invariance, we can use this result in both Minkowski
and Euclidean space to identify and isolate the ultraviolet pole. It
is now straightforward to show two important results that follow
from the antisymmetries built into Eq.\ (\ref{gggprime}).

First, working in Minkowski space, we can readily confirm the
vanishing of $F_{3g}$ for arbitrary massless $\beta_i$.   In this
case, the function remains scale-invariant in the $\beta_i$,
although of course we cannot rescale by $\beta_i^2=0$.
Nevertheless,  the explicit form of $g(\zeta,\vep)$ is
\bea g(\zeta,\vep)=\frac{1}{2\vep}\frac{1}{\zeta}\quad \quad
(\beta^2=0)\, , 
\label{gzero}
\eea
which, using $\zeta(dg/d\zeta)=-g$, immediately gives a vanishing
integrand in Eq.\ (\ref{gggprime}) by antisymmetry. It is
interesting to note that, unlike the discussion in momentum space in
Ref.\ \cite{Aybat}, this proof of the vanishing of the three-gluon
diagram, $F_{3g}^{(2)}$, Fig.\ \ref{diagrams}a, does not require a
change of variables.

In fact, the vanishing of Fig.\ \ref{diagrams}a
extends to the case where only two of the
three lines are massless \cite{Gardi}.   Taking for definiteness $\beta_1^2=\beta_2^2=0$
with $\beta_3^2\ne 0$,
and using Eq.\ (\ref{gzero}) in Eq.\ (\ref{gggprime}), 
we find
\bea F^{(2)}_{3g}(\beta_I,\varepsilon) &=& -\ \int d^Dx \,
 \frac{1}{\left(\sqrt{x^2}\right)^{4-6\vep}}
\nonumber\\
&\ & \hspace{-20mm}
\times\, \left( g(\zeta_3,\vep) + \zeta_3 \frac{\partial g(\zeta_3,\vep)}{\partial\zeta_3}\right)
\left( \frac{\beta_1\cdot\beta_3}{\zeta_1}- \frac{\beta_2\cdot\beta_3}{\zeta_2}\right)
 \,. \nonumber\\
 \label{gzeroF}
 \eea
 In this case, we follow \cite{Aybat} and make a change of variables,
 using the light-like directions $\beta_{1,2}$ to define light-cone coordinates.
To be specific, if we choose $\chi_1=\zeta_1/\beta_1\cdot\beta_3$
 and $\chi_2=\zeta_2/\beta_2\cdot\beta_3$, we derive an 
 integrand that is manifestly antisymmetric in $\chi_1$ and $\chi_2$.
We  note that the momentum
space method of \cite{Aybat} also applies directly to the case when two of the
three Wilson lines are massless, although
this was not pointed out explicitly there.   
In both cases, the relevant change of variables exchanges
two lightlike directions.
This approach does not
show that diagrams with a single massless line vanish identically,
and indeed this seems unlikely, given that in this case there
is only a single lightlike direction.

Our second main result is that for both massive and massless Wilson lines
the function $\Gamma^{(2)}_{3g}(\beta_i\cdot\beta_j)$ vanishes when
any pair of the invariants are equal, say,
$\beta_1\cdot\beta_2=\beta_1\cdot\beta_3$. We can show this by
changing variables in Euclidean space from $x_i$ to $r=\sqrt{x^2}$
and the $\zeta_i=\beta_i\cdot x/\sqrt{x^2}$.   A straightforward
calculation gives a form in which the overall scalelessness of the
diagram is manifest in the radial integral,
\bea F^{(2)}_{3g}(\beta_I,\varepsilon) &=& -\ \int
\frac{dr}{r^{1-4\vep}} \, \int \frac{d\zeta_1 d\zeta_2
d\zeta_3}{\sqrt{K(\zeta_I)}}
\nonumber\\
&\ & \hspace{-25mm} \times
\sum_{i,j,k=1}^3
\ep_{ijk} \,
\zeta_k\zeta_i\,\beta_i\cdot \beta_j
\ g(\zeta_j,\vep) g(\zeta_k,\vep)\
 \frac{ \partial g(\zeta_i,\vep)}{\partial \zeta_i}\, .
 \label{gggprime2}
\eea
Defining $w_j\equiv \beta_1\cdot\beta_j$, $\xi_j\equiv
\zeta_j-w_j\zeta_1$, and
$z_3\equiv[\beta_2\cdot\beta_3-w_2w_3]/\sqrt{1-w_2^2}$, the explicit
form of $K(\zeta_I)$ for arbitrary $w_2$ and $w_3$ is
\bea K(\zeta_I) &=& - (1-w_2^2)\xi_3^2 - (1-w_3^2)\xi_2^2
+\ 2\xi_2\xi_3z_3\sqrt{1-w_2^2} 
\nonumber\\
 &\ & \hspace{5mm} +
(1-\zeta_1^2)(1-w_2^2)(1-z_3^2-w_3^2)\, .
\label{Keye}
\eea
The function $K(\zeta_I)$ is symmetric under the interchange of
$\zeta_2$ and $\zeta_3$ when $w_2=w_3$. We easily check that when
$w_2=w_3$ the remaining integrand in Eq.\ (\ref{gggprime2}) is
antisymmetric under the exchange of $\zeta_2$ and $\zeta_3$.   The variables
$\zeta_2$ and $\zeta_3$ are exchanged by a reflection in the two-sphere
defined by any fixed value of $\zeta_1$ about the axis specified by
  the projection of $\beta_2+\beta_3$ into this two-sphere.
 The integration region of $\zeta_2$ and $\zeta_3$ is therefore
 also symmetric.   We conclude that the integral over 
  $\zeta_2$ and $\zeta_3$ in Eq.\ (\ref{gggprime2}) 
vanishes  when $\beta_1\cdot\beta_2 =\beta_1\cdot
\beta_3$.    This relation holds as well in Minkowski space, which
can be reached by analytic continuation at fixed ratios of the inner
products $\beta_1\cdot \beta_2/\beta_1\cdot \beta_3$.

\begin{figure}[t]
\centerline{\epsfxsize=7cm \epsffile{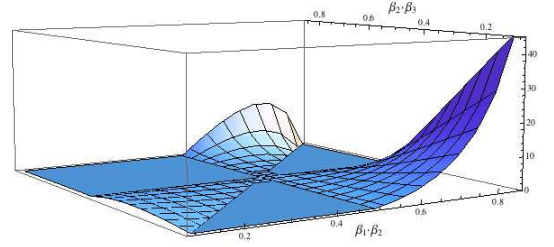}}
\caption{\label{plotfig} Plot of the integral in (\ref{gggprime})
for $\beta_1\cdot\beta_3=0.5$.}
\end{figure}

We do not have here an analytic form for the residue of the UV pole
of $F^{(2)}_{3g}$ for generic values of the invariants
$\beta_i\cdot\beta_j$. A numerical analysis of Eq.\
(\ref{gggprime}), however, is particularly straightforward in
Euclidean space. For this purpose, it is convenient to use
$D$-dimensional polar coordinates, $r,\, \Omega_{D-1}$. The single
overall ultraviolet pole in the scaleless integral (\ref{gggprime})
appears at $r=0$, and the remaining three angular integrals at
$\vep=0$ determine the residue of the pole in $\overline{\rm MS}$
renormalization. These can be carried out readily, using the
elementary form of the function $g(\zeta,0)$ in Eq.\ (\ref{fdefg}),
\bea g(\zeta,0) = \frac{\pi - \arccos{\zeta}}{\sqrt{1-\zeta^2}}\, .
\eea In Fig.\ \ref{plotfig}, we show a plot of the residue of the UV
pole of $F^{(2)}_{3g}$ for $\beta_1\cdot\beta_3=0.5$ in the
$\beta_1\cdot\beta_2/\beta_2\cdot\beta_3$-plane, suppressing the
surface where the integral is negative for clarity.   Notice the
lines of zeros at $\beta_1\cdot\beta_2 = \beta_2\cdot\beta_3$ and at
$\beta_1\cdot\beta_2=0.5$ and $\beta_2\cdot\beta_3=0.5$. The peak
towards $\beta_1\cdot\beta_2\rightarrow 1$ reflects an additional
singularity in the integrand when two velocities are parallel.

We now turn to the remaining 3E ``double-exchange" diagrams, Fig.\
1b,c. These diagrams are given by four $\lambda$ integrations, two
of which are along the $\beta_j$ line. For example, Fig.\  1b may be
written in the notation introduced above as
\bea M_{b} &=& C_{1b}\; {\cal {\overline N}}(\vep)\;
(\beta_i\cdot\beta_j)\; (\beta_j\cdot\beta_k) \int_0^\infty
d\lambda_{j,a} \int_0^{\lambda_{j,a}} d\lambda_{j,b}
\nonumber\\
&\ & \hspace{10mm} \times\, f(\lambda_{j,a}\beta_j,\beta_i,\vep)\,
f(\lambda_{j,b}\beta_j,\beta_k,\vep)\, , 
\label{Msubb}
\eea
where the functions $f$ are defined as in Eq.\ (\ref{fdefg}) with
$x=\lambda_{j,c}\beta_j$ and with $c=a,b$.
${\cal {\overline N}}(\vep)$ absorbs overall factors that are finite
in the limit $\vep = 0$. In the following we set ${\cal {\overline
N}}(\vep) = 1$. We have kept the overall color factor, represented
by $C_{1b}$. The variables $\zeta_i=\beta_i\cdot x/\sqrt{x^2}$,
are independent of the scale of $x$ in Eq.\ 
(\ref{Msubb}), so that all
dependence on the $\lambda_{j,c}$ variable is in the overall factor
of $x^2=\lambda_{j,c}^2$, giving
\bea M_{b} &=& C_{1b}\; (\beta_i\cdot\beta_j)\;
(\beta_j\cdot\beta_k)\; \int_0^\infty d\lambda_{j,a}
\int_0^{\lambda_{j,a}} d\lambda_{j,b}\
\nonumber\\
&\ & \hspace{10mm} \times\, \frac{g(\beta_i\cdot\beta_j,\vep)\,
g(\beta_k\cdot\beta_j,\vep)}{
\left(\lambda_{j,a}\lambda_{j,b}\right)^{1-2\vep}} \, . \eea
The exchange diagrams contribute to two color structures, symmetric
and antisymmetric with respect to the generators $T_a$ and $T_b$ on
line $\beta_j$ that are linked by gluon exchange to the lines
$\beta_i$ and $\beta_k$, respectively.  Combining Fig.\
\ref{diagrams}b and c, we thus write
\bea M_{b} (\beta_I,\vep) + M_{c} (\beta_I,\vep)=
M_{b+c}^{(A)}(\beta_I,\vep) \ +\
M_{b+c}^{(S)}(\beta_I,\vep) \, .\nonumber\\
\eea
The first, antisymmetric, combination has the same color factor as
the three gluon diagram, Fig.\ \ref{diagrams}a, discussed above.
It is given by
\begin{widetext}
\bea M_{b+c} ^{(A)}(\beta_I,\vep) &=& \frac{1}{2} \left[ C_{1b}\ -\
C_{1c} \right]\; (\beta_i\cdot\beta_j)\; (\beta_j\cdot\beta_k)\;
\left(\int_0^\infty d\lambda_{j,a} \int_0^{\lambda_{j,a}}
d\lambda_{j,b} - \int_0^\infty d\lambda_{j,b} \int_0^{\lambda_{j,b}}
d\lambda_{j,a} \right) \nonumber\\
&\ & \hspace{10mm} \times\, \frac{g(\beta_i\cdot\beta_j,\vep)\,
g(\beta_k\cdot\beta_j,\vep)}{
\left(\lambda_{j,a}\lambda_{j,b}\right)^{1-2\vep}} \, ,
\eea
\end{widetext}
where the difference in color factors in square brackets produces a
commutator for the color matrices on the $\beta_j$ line. Clearly,
the integrals in parenthesis are identical and cancel. 
This result applies to arbitrary masses for the Wilson lines.
As in the
massless case, 
only the symmetric contribution of the
double exchange diagrams survives and contributes to the soft
 matrix.  As described in Ref.\ \cite{Aybat}, however, the
symmetric contribution to the soft matrix 
at two loops
is already generated by
the exponential of the one-loop anomalous dimension in Eq.\
(\ref{expoS}).

The reasoning above has a significant phenomenological application
to $2\rightarrow 2$ production processes. Using Wick rotation, the
vanishing of 3E diagrams after one-loop renormalization when pairs
of invariants are equal applies to $2 \rightarrow 2$ processes
involving the production of pairs of heavy quarks from light quarks
or gluons. In particular, we consider Wilson line velocities
corresponding to momentum configurations with
$t=(p_1-p_3)^2=(p_1-p_4)^2=u$ in Eq.\ (\ref{f_1f_n}) with $n=2$
\cite{footnote}. Note that because it is trivial to reintroduce
$\beta_i^2$-dependence,  the result applies as well to the limit
where one or more line becomes massless.

\begin{figure}
{\epsfxsize=2.5 cm \epsffile{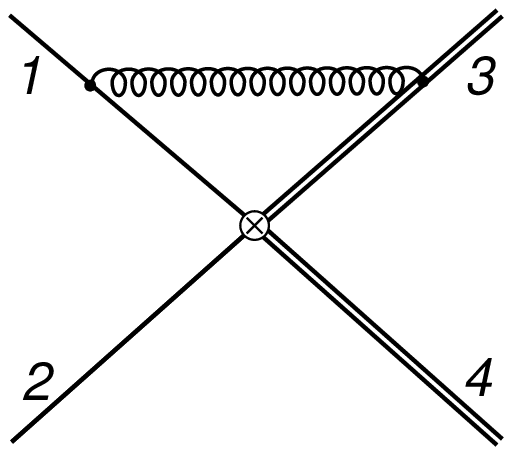} \quad \epsfxsize=2.5 cm \epsffile{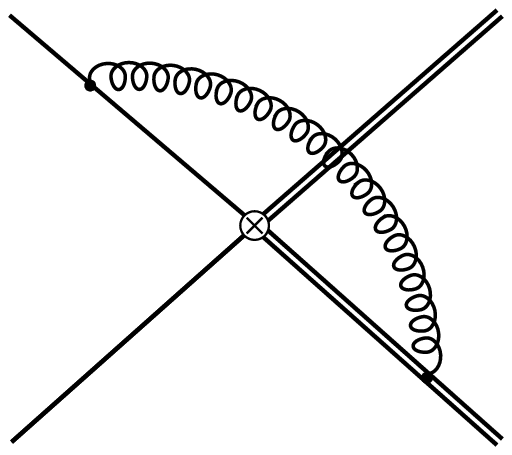} \\
\hbox{ \hskip 2.25 cm (a) \hskip 3.0 cm (b) }
\quad \epsfxsize=2.5 cm \epsffile{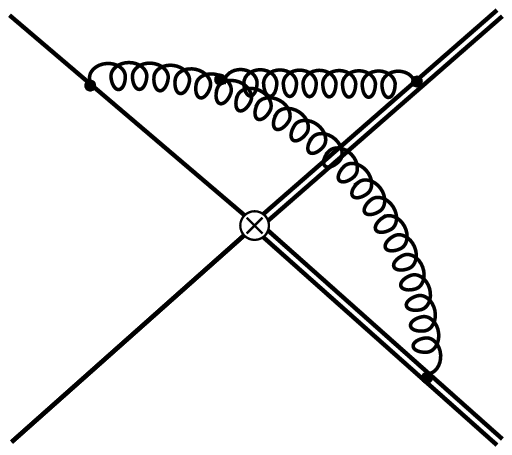} \quad \epsfxsize=2.5 cm \epsffile{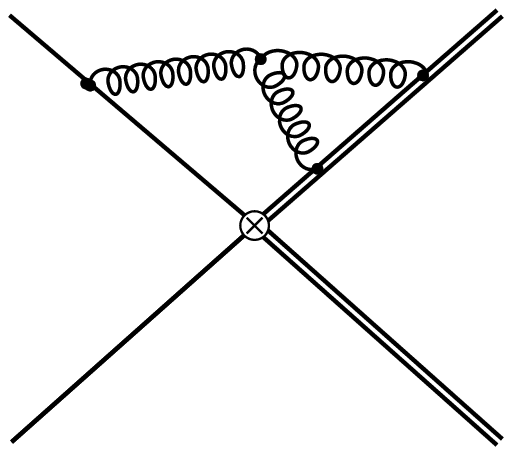}\\
\hbox{ \hskip 2.25 cm (c) \hskip 3.0 cm (d) }
\quad \epsfxsize=2.5 cm \epsffile{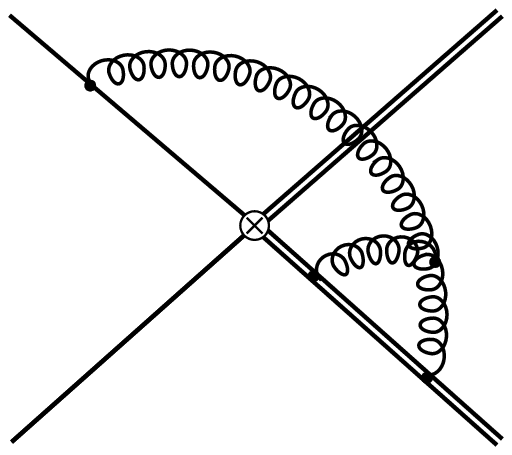} \quad \epsfxsize=2.5 cm \epsffile{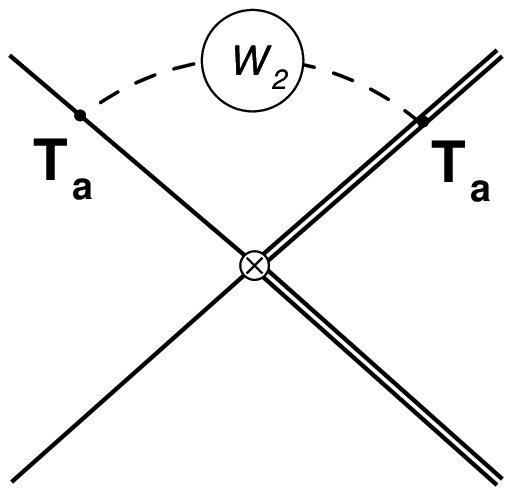}\\
\hbox{ \hskip 2.25 cm (e)  \hskip 3.0 cm (f)}}
\caption{ One-loop 2E diagrams   (a,b) that cancel at
$u=t$. (c) representative 3E diagram for this process.
Diagrams (d,e) are two-loop diagrams that cancel.   Diagram
f represents the color structure of the webs. \label{2Ediagrams}}
\end{figure}

Of special interest are the anomalous dimension matrices that enter
pair production: the $2\times 2$ matrix for $q\bar{q} \to Q\bar{Q}$
and the $3\times 3$ matrix for $gg \to Q\bar{Q}$. At one loop, these
matrices are diagonal in their $s$-channel singlet-octet bases at
$u=t$ \cite{Kidonakis:1997gm}. Specifically, the off-diagonal
elements of these $\Gamma_S^{(1)}$ are all proportional to
$\ln(u/t)$. One- and two-loop diagrams contributing to these
anomalous dimension matrices are illustrated by Fig.\
\ref{2Ediagrams}, where the single eikonal lines represent incoming
light partons (quark pairs or gluons) and the double lines heavy
quark pairs. At $u=t$, the off-diagonal zeros in $\Gamma_S^{(1)}$
directly reflect cancellation between pairs of diagrams such as
those in Fig.\ \ref{2Ediagrams}a and b.

To illustrate the pattern of cancellation, we continue the notation
above and write the amplitude, $M_D$ corresponding to eikonal
diagram $D$ as the product of a color factor and a
velocity-dependent factor,
\bea M_D^{(n)} = C_D\, F_D^{(n)}(\beta_I)\, .
 \label{MD}
\eea
In these terms, the vanishing of matrix elements that describe the
mixing of color singlet-octet tensors in  $\Gamma_S^{(1)}$ follows
from identities relating the two diagrams,
\bea F^{(1)}_{2a}(\beta_I) &=& F^{(1)}_{2b}(\beta_I)\, ,
\label{Fplus}\\
C_{2a} &=&-C_{2b}\, ,
\label{Cminus}
\eea
which are easily verified at one loop for $u=t$. We have shown above
that at two loops with $u=t$, 3E diagrams like Fig.\
\ref{2Ediagrams}c vanish independently of their color structure,
leaving only 2E diagrams and the color-symmetric parts of
3E exchange diagrams at two loops.      This suggests that at two loops the
same pattern of cancellation may persist for $\Gamma_S^{(2)}$. This
is indeed the case, as we now explain.

At one loop, the expansion of Eq.\ (\ref{expoS}) in  terms of
one-loop anomalous dimensions can be thought of as the sum of
contributions from each gluon-exchange diagram. Expanding the
one-loop form of Eq.\ (\ref{expoS}) to order $\alpha_s^2$, we
generate all color-symmetric contributions of two-loop ladder
diagrams, corresponding for example to the color-symmetric
contributions of Fig.\ \ref{diagrams}d and e above. The sum of all
2E diagrams is thus easily rewritten in terms of the expansion of
Eq.\ (\ref{expoS}) in terms of $\Gamma_S^{(1)}$, plus  two-loop
contributions that involve commutators of generators on one or more
of the Wilson lines.

Examples of diagrams with antisymmetric color structure, which do
not correspond to the expansion of the one-loop anomalous
dimensions, are Figs.\ \ref{diagrams}f and \ref{2Ediagrams}d and e,
and the color-antisymmetric part of Fig.\ \ref{diagrams}e. These
diagrams contribute only their antisymmetric combinations of color
generators to $\Gamma_S^{(2)}$, and are thus proportional to $C_A$
times a one-loop color factor. Figure \ref{2Ediagrams}f is a
schematic representation of the color structure of such diagrams and
of their contributions to the two-loop anomalous dimension matrices.

We note that exactly as for the massless case
in Ref.\ \cite{Aybat}, all 2E diagrams are 
generated from an exponential of  ``webs" \cite{webs}. Webs are 
themselves 2E diagrams that are
two-particle irreducible under cuts of the Wilson lines, starting
with single-gluon exchange. The contributions of two-loop webs are
precisely the antisymmetric color structures that we have just
identified.

As we have observed, Eq.\ (\ref{Cminus}) holds for the color factors
of Fig.\ \ref{2Ediagrams}a and b.   The same relation then
holds for the web contributions of Fig\ \ref{2Ediagrams}d and e,
because they are proportional to the same color structure as the
one-loop diagrams, as illustrated in Fig.\ \ref{2Ediagrams}f.
Therefore, whenever the identity for velocity factors, Eq.\ (\ref{Fplus}),
holds at two loops, this pair of two-loop diagrams cancels. The 2E
diagrams of Fig.\ \ref{2Ediagrams}a and b, however, can depend only
on the invariants formed from the two eikonal velocities in
question, so that at $u=t$, 
\bea F^{(2)}_{2a}(\beta_I) =
F^{(2)}_{2b}(\beta_I) \quad (u=t)\,\, . 
\eea 
This relation will hold as well for
diagrams associated with crossed ladders, as in Fig.\
\ref{diagrams}e, and other diagrams with color-antisymmetric
contributions, including those with self-energies of the exchanged
gluon and vertex corrections on the massive eikonal. As a result,
cancellations between pairs of 2E diagrams that occur at one loop
recur at two.
For $2\rightarrow 2$ kinematic configurations with $u=t$, then,
we have in place of Eq.\ (\ref{Gamma2}) the relation
\bea {\bf \Gamma}_{S_\f}^{(2)} (\beta_I) = D(\beta_I)\, {\bf
\Gamma}_{S_\f}^{(1)}(\beta_I) \quad (u=t)\, , 
\label{Gamma2m}
\eea
where $D(\beta_I)$ is a matrix that is diagonal in the $s$-channel
singlet-octet basis. For massless two-loop cases, the matrix
$D(\beta_I)$ is proportional to the identity, and we recover a
special case of Eq.\ (\ref{Gamma2}). More generally, however, the
integrals of the 2E webs depend on masses.   We will give explicit
results elsewhere. In any case, the relation (\ref{Gamma2m}) applies
for scattering at ninety degrees in the center of mass, including
the limiting case of production at rest ($\hat{s}=4m_Q^2$), of
particular relevance to threshold resummation for the total cross
section for heavy quark production \cite{Heavyquarktotal}.
A subset of these diagrams have been analyzed very recently in \cite{Kidonakis:2009ev}.

For massless particles, as noted in \cite{Aybat} it is certainly
natural to generalize the result Eq.\ (\ref{Gamma2}) for lightlike
partons to all orders.   This possibility has been explored recently
in Refs.\ \cite{BecherNeubertGardiMagnea}. The result that we have
found here, that Eq.\ (\ref{Gamma2}) does not apply to arbitrary
massive kinematics, suggests that the zero-mass case is indeed
special.   We believe this should encourage the search for a
symmetry or principle underlying the relation, Eq.\ (\ref{Gamma2}).

In summary, we have shown that for products of massive Wilson lines, 
one- and two-loop soft anomalous dimensions are generally not proportional.
We have noted, however, that diagrams that link two massless with
one massive line through the three-gluon coupling
vanish by a variant of the reasoning for three massless lines.
We have also shown that diagrams that link a single massless line with
two massive lines cancel for the case of two-to-two production processes
at threshold, and when $u=t$, that is, for production at ninety degrees in the
center of mass system.   For these momentum configurations,
the two-loop anomalous dimension matrix is diagonal in the same
color basis as the one loop, although they are not related by a 
simple proportionality as in the massless case.
The color-antisymmetric parts of exchange diagrams linking
three eikonal lines cancel independently of masses.
We carried out a numerical evaluation of the three-gluon diagram in Euclidean space,
but the qualitative conclusions of this paper extend to Minkowski space.
Whatever the field-theoretic origin of the results described above
for massive and massless partons, the renormalization of the
composite operators that link Wilson lines will be relevant to the
analysis of light-parton jets, heavy quarks and potential new,
strongly interacting particles at the Tevatron and the LHC.

\acknowledgments
This work was supported by the National Science
Foundation, grants PHY-0354776, PHY-0354822 and PHY-0653342.  The
work of AM is supported by a fellowship from the {\it US LHC Theory
Initiative} through NSF grant 0653342.  We thank Michal Czakon, Lance Dixon, Einan
Gardi and Lorenzo Magnea for helpful exchanges.


\begin{thebibliography}{}
\bibitem{footnote}
We observe that the relations $s=-t$ or $s=-u$ do not correspond to
equal invariants in the sense above.
\end{thebibliography}
\end{document}